\documentstyle[aps,preprint]{revtex}
\begin{document}
\draft
\tighten
%\twocolumn[\hsize\textwidth\columnwidth\hsize\csname @twocolumnfalse\endcsname
\preprint{\vbox{
                \hfill TIT/HEP-413/NP
 }}
\title{
Comment on ``Determination of pion-baryon coupling constants\\ 
from QCD sum rules"}
\author{Hungchong Kim\footnote{Email : hckim@th.phys.titech.ac.jp}}
\address{Department of Physics, Tokyo Institute of Technology, Tokyo 152-8551,
Japan\\}
\maketitle

\begin{abstract}

In this comment, we propose possible errors in constructing the continuum
contribution in the sum rule studied by
M. C. Birse and B. Krippa, Phys. Rev. C. {\bf 54} (1996) 3240.   
\end{abstract}

%\vspace{30pt}

In Ref.~\cite{krippa}, Birse and Krippa [BK] have 
calculated $\pi NN$, $\pi \Sigma \Sigma$, and
$\pi \Sigma \Lambda$ coupling constants using  QCD sum rules. 
In this comment, we want to point out that 
they might not treat the continuum contribution properly.
If  the continuum contribution is treated  properly within
their analysis, the results provided in Ref.~\cite{krippa} need to be 
significantly modified.

BK considered the two-point correlation function
\begin{eqnarray}
\Pi (p) = i \int d^4 x e^{i p \cdot x} \langle 0 | T[\eta_p (x)
{\bar \eta}_n (0)]| \pi^+ (k) \rangle\ .
\label{two2}
\end{eqnarray}
Here $\eta_p$ is the proton interpolating field of Ioffe~\cite{ioffe} and
$\eta_n$ is the neutron interpolating field. For
$\pi \Sigma \Sigma$ and $\pi \Sigma \Lambda$ couplings, 
they use the interpolating fields for $\Sigma^{+,0}$ and $\Lambda$
obtained from the nucleon interpolating field by using SU(3) rotation.
They constructed the sum rules in the leading order of the pion
momentum $k$ and considered the structure $\not\!k \gamma_5$.

In the conventional QCD sum rules, the continuum contribution is modeled 
such that its spectral density is given by 
a step function which starts from a threshold, say $S_{\pi}$.
According to the duality argument, the coefficient of the step function is 
obtained from 
the terms containing $ln (-p^2)$ in the OPE. 
More specifically, the spectral density for the continuum is
$\rho_c^{phen}
\equiv \rho^{ope} (p^2)  \theta(p^2 -S_{\pi})$, where $\rho^{ope}$ is
the spectral density of the OPE. 
 
The sum rule equation after the Borel transformation is given by
\begin{eqnarray}
\int^\infty_0 ds e^{-s/M^2} [\rho^{ope}(s) - \rho^{phen} (s)] =0\ ,
\label{sum1}
\end{eqnarray}
where $M$ is the Borel mass. Within this approach, it is straightforward
to construct the sum rule equations. Then it is easy to prove
that $E_2$, $E_1$, and $E_0$ factors appearing in
Eqs.(17),(28),(29) and in the numerator of Eq.(36) of Ref.~\cite{krippa} 
should be replaced as follows,
\begin{eqnarray}
E_2 \rightarrow E_1\;; \quad E_1 \rightarrow E_0\;; \quad 
E_0 \rightarrow 1 \ .
\end{eqnarray}

To be more specific, let's consider the second term in Eq.(17) of
Ref.~\cite{krippa} and
prove the replacement of  $E_1 \rightarrow E_0$.
This term comes from Eq.(13) whose form can be written as
$ C~ {\rm ln} (-p^2)$ where $C$ is a constant.
The corresponding spectral density is $ C~ \theta(p^2)$
which provides the continuum spectral density 
as $ C~ \theta(p^2 -S_{\pi})$.
Then the integration over $p^2$ with the Borel weight,
\begin{eqnarray}
\int^\infty_0 dp^2  e^{-p^2/ M^2} C \theta(p^2 -S_{\pi N})
=
C~M^2 e^{-S_{\pi}/M^2} \ .
\end{eqnarray}
And the corresponding integration for the OPE part yields  $C M^2$.
By moving the continuum part to the OPE side, one obtains the term
\begin{eqnarray}
C M^2 (1 - e^{-S_{\pi}/M^2})\ . 
\end{eqnarray}
Note that the factor $(1 - e^{-S_{\pi}/M^2})$ 
is the definition of $E_0$, not $E_1$ !  This $E_0$ factor should appear
in the second term of Eq.(17) instead of $E_1$.  
The third replacement, $E_0 \rightarrow 1$, is easy to see because this 
$E_0$ factor in Ref.~\cite{krippa} comes from the OPE term of $1/p^2$. 
Obviously, the spectral density of $1/p^2$ is just a delta function, 
$\delta(p^2)$, and it can not contribute to the continuum. 
Similarly, one can derive the replacement, $E_2 \rightarrow E_1$.

Then how these replacements affect their results ?
In figure 1 of Ref.~\cite{krippa}, 
after these replacements, the dashed line now varies from 13 to 30 
the solid line varies from 7.5 to 5. Therefore, the curves one gets
after the replacements
are  clearly different from what BK provided in Ref.~\cite{krippa},
which of course changes their results substantially.  Additional
discussion on this sum rule using the correct continuum  
can be found in Ref.~\cite{hung}.

According to Ref.~\cite{birse}\footnote{We thank M. C. Birse for
clarifying the steps leading to the factors in question}, 
it is claimed that the use of a double dispersion relation~\cite{ioffe2} 
is crucial in deriving those continuum factors in question.
Within a double dispersion relation, the perturbative spectral
density is claimed to be of the form, 
\begin{eqnarray}
\rho(s_1, s_2) = b(s_1) \delta(s_1 -s_2)\ .
\label{dspe}
\end{eqnarray}
%Within the method provided by Ref.~\cite{birse}, $E_1$ factor can be
%obtained from the OPE term of $C ln(-p^2)$.   
%First, 
The function $b(s)$ in Eq.(\ref{dspe}) is determined via 
the relation, 
\begin{eqnarray}
\int_0^\infty ds_1 \int_0^\infty ds_2 {\rho(s_1, s_2) \over 
(s_1 - p^2) (s_2 -p^2)} &=& \int_0^\infty {b(s) \over (s-p^2)^2}\nonumber\\
&=& C ln(-p^2)
\end{eqnarray}
The second relation is satisfied for $b(s) = -C s$, up to some
subtraction terms. 

According to Ref.~\cite{birse}, the continuum contribution 
from $C ln (-p^2)$ is obtained by
putting Eq.(\ref{dspe}) (with $b(s) = -C s$) into the double
dispersion integral,
\begin{eqnarray}
\Pi (p^2) \equiv \int_{S_\pi}^\infty ds_1 \int_{S_\pi}^\infty 
ds_2 {\rho(s_1, s_2) \over
(s_1 - p^2) (s_2 -p^2)} &=& - \int_{S_\pi}^\infty {C s \over 
(s-p^2)^2}\nonumber\\
&=& C \Big [ ln (S_{\pi} - p^2) - {S_{\pi} \over S_{\pi} - p^2}
+ \cdot \cdot \cdot \Big]
\label{bcont} 
\end{eqnarray}
Then BK performed  a single Borel transformation, transfered this
continuum to the OPE side, and obtained the factor, $E_1$.
Similarly, one can get the factors, $E_0$ and $E_2$, by following the
similar steps.

However, in this derivation, it is important to note that
the term, $S_{\pi} /  (S_{\pi} - p^2)$,
in Eq.~(\ref{bcont})
is crucial in obtaining the $E_1$ factor.  Without this, one would
obtain $E_0$, the same factor that  we have derived using the conventional 
method mentioned above. 
The situation is similar in producing the factors,  $E_0$ and $E_2$,
in BK sum rule.
What is the physical meaning of this new term ?  If one takes the  
discontinuity of the correlator, 
$\Delta \Pi \equiv  \Pi(p^2+i \epsilon) - \Pi(p^2 - i \epsilon)$,
then the new term yields a pole at $p^2 = S_\pi$ in addition to a
step-like continuum coming from the term $ln (S_{\pi} - p^2)$.
The step-like continuum is well understood from the QCD duality
but the pole at the continuum threshold does not have any physical
meaning.  This pole is  produced by the mathematical way of constructing
the continuum and therefore needs to be subtracted out.
Otherwise, the duality of QCD in constructing the continuum in QCD
sum rule can not be satisfied by the presence of this pole at
the continuum threshold.
Once the pole subtracted out, then the factors in question can
be replaced as we suggested.

\acknowledgments
This work is supported by Research Fellowships of the Japan Society for the
Promotion of Science.

\end{document}